\documentclass[10pt]{article}
\pdfoutput=1
\usepackage{cite}
\usepackage{amsmath,bbm}
\usepackage{amssymb}
\usepackage{graphicx} \graphicspath{{images/}}
\usepackage[latin1]{inputenc}
\usepackage{mathrsfs}
\usepackage{mathtools}
\usepackage[dvipsnames]{xcolor}
\usepackage[footnotesize]{caption}
\usepackage{xcolor}
\usepackage{hyperref}
\usepackage[hmarginratio=1:1,top=32mm,columnsep=20pt]{geometry}
\usepackage{paralist}
\usepackage{abstract}
\usepackage{upgreek}
\usepackage{cancel}
\usepackage{multirow}
\usepackage[normalem]{ulem}

\newcommand{\fsl}[1]{\ensuremath{\mathrlap{\!\not{\phantom{#1}}}#1}}

\makeatletter
\let\@fnsymbol\@arabic
\makeatother

\newcommand{\bea}{\begin{eqnarray}}
	\newcommand{\eea}{\end{eqnarray}}
\newcommand{\be}{\begin{equation}}
	\newcommand{\ee}{\end{equation}}
\newcommand{\ba}{\begin{array}}
	\newcommand{\ea}{\end{array}}

\def\gsim{\mathrel{\rlap{\lower4pt\hbox{\hskip1pt$\sim$}}
		\raise1pt\hbox{$>$}}}

\begin{document}
	
	\thispagestyle{empty}
	\begin{flushright}
		APCTP Pre2019-023\\
		\vspace*{2.mm} \today
	\end{flushright}
	
	\begin{center}
		{\Large \textbf{
				Searching for Dark Photons at the LHeC and FCC-he
			}
		}  
		
		\vspace{0.5cm}
		Monica~D'Onofrio$^\star$\footnote{Monica.D'Onofrio@cern.ch}
		Oliver~Fischer$^\dagger$\footnote{oliver.fischer@kit.edu}
		and	Zeren~Simon~Wang$^{\circ\ddagger}$\footnote{zerensimon.wang@apctp.org}
		
		\medskip
		
		{\small \textit{ 
				${}^\star$ 
				Department of Physics, University of Liverpool, \\ 
				\normalsize Oliver Lodge, Oxford Street, Liverpool L69 7ZE, UK\\[0pt] 
		}}    
		{\small \textit{ 
				${}^\dagger$ 
				Institute for Nuclear Physics, Karlsruhe Institute of Technology, \\
				\normalsize	Hermann-von-Helmholtz-Platz 1, D-76344 Eggenstein-Leopoldshafen, Germany\\[0pt] 
		}}
		{\small \textit{ 
				${}^\circ$ 
				Physikalisches Institut der Universit\"at Bonn, Bethe Center for Theoretical Physics, \\ 
				\normalsize Nu{\ss}allee 12, 53115 Bonn, Germany\\[0pt] 
		}}  
		
		{\small \textit{ 
				${}^\ddagger$ 
				Asia Pacific Center for Theoretical Physics (APCTP) - Headquarters San 31,\\ 
				\normalsize Hyoja-dong, Nam-gu, Pohang 790-784, Korea\\[0pt] 
		}}

		\setcounter{footnote}{0}
		
		\vspace*{0.7cm}
		
		\begin{abstract}
			Extensions of the Standard Model (SM) gauge group with a new $U(1)_X$ predict an additional gauge boson. Through kinetic mixing with the SM photons featured by a coupling $\epsilon$, the ensuing so-called dark photons $\gamma'$, which acquire mass as a result of the breaking of the gauge group $U(1)_X$, can interact with the SM field content. These massive dark photons can therefore decay to pairs of leptons, hadrons, or quarks, depending on their mass $m_{\gamma'}$. In this work, we discuss searches for dark photons in the mass range around and below one GeV at the LHeC and FCC-he colliders. The signal is given by the displaced decays of the long-lived dark photon into two charged fermions. We discuss the impact of conceivable irreducible (SM and machine-related) backgrounds and different signal efficiencies. Our estimates show that the LHeC and FCC-he can test a domain that is complementary to other present and planned experiments. 
		\end{abstract}
		
	\end{center}

	\newpage
	\section{Introduction}
	\label{sec:intro}
	In the class of hidden sector theories, new particles are predicted to interact with the Standard Model (SM) field content via feebly coupled mediator particles. This class can be categorized via a small number of so-called ``portals'': the scalar, pseudoscalar, vector, and neutrino portal, within which the interacting mediator particle is given by a second Higgs boson, axion-like particles, dark photons ($\gamma'$) and heavy neutral leptons, respectively.
	In the case of the vector portal, a dark photon, kinetically mixed with the SM photon, is predicted as the gauge boson of an extra gauge group $U(1)_X$ \cite{Okun:1982xi,Galison:1983pa,Holdom:1985ag,Boehm:2003hm,Pospelov:2008zw}. Through for instance a SM-like Higgs mechanism, this hidden $U(1)_X$ gauge group is spontaneously broken, giving rise to a massive dark photon. As a result of the kinetic mixing, dark photons are coupled to the SM electromagnetic current in the same way as the SM photons are, except that this coupling is suppressed by a small mixing angle. Another consequence of this mixing is that the SM quarks and charged leptons acquire a milli-charge under the new gauge group. 
	Depending on their mass, dark photons have different decay channels.
	For a mass smaller than $2 m_e$ the only viable decay channel is into three photons. For larger masses, as considered in this work, decays to pairs of charged leptons, quarks, or mesons become kinematically allowed and are dominant.
	
	It is important to notice that the dark photon models presently studied in the literature can be categorized loosely into two classes: minimal and non-minimal.
	Models from the former class, so-called minimal models, consider only gauge mixing and dark photons can be produced from decays of light mesons such as the pion and $\eta$, from Bremsstrahlung processes, or directly from interactions of particle beams at colliders. 
	An eminent example for models from the second class, so-called non-minimal models, is Dark Supersymmetry, where the dominant hidden-sector interactions with the SM can be given via a Higgs portal and the dominant dark photon production can come from Beyond-the-Standard-Model (BSM) Higgs decays. 
	In general the non-minimal setups feature additional particles that can participate in production and decay of the dark photon, and search prospects depend hence strongly on the model assumptions. In this work, we focus on the minimal scenario.
	
	The most stringent existing limits in the low-mass and low-coupling region are combinations from dark photon searches at the beam dump experiments E141 \cite{Riordan:1987aw}, E774 \cite{Bross:1989mp}, and one in Orsay \cite{Davier:1989wz}; the strongest constraints in the low-mass ``large''-coupling regime (``large'' denoting mixings of at least $10^{-3}$) stem from the beam-dump experiment NA48 \cite{Batley:2015lha} and the electron-positron collider experiment BaBar \cite{Lees:2014xha}. 
	For a detailed list of experimental searches for dark photons see for instance Ref.~\cite{Bauer:2018onh} and references therein. In Ref.~\cite{Ilten:2018crw}, both a summary of present and future bounds on dark photons and a recasting tool for different dark photon models are provided.

	Dark photons are also being searched for in collider experiments, with LHCb targeting minimal models whilst ATLAS and CMS are mostly considering non-minimal ones. The CMS collaboration at the LHC searches for displaced dark photons in the context of Dark Supersymmetry models \cite{CMS:2018rdr, CMS:2018lqx} requiring dark photon masses above the di-muon threshold since the triggering is via the muons. The ATLAS experiment searches for prompt and displaced lepton-jets of dark photons in SUSY-portal and Higgs-portal models \cite{Aad:2015sms,Aad:2014yea,ATLAS:2016jza} considering both electron and muon final states. The LHCb experiment searches for both prompt-like and long-lived dark photons focusing on $\gamma' \rightarrow \mu\mu$ decays \cite{Aaij:2017rft} and is also only sensitive to dark photon mass larger than $2 m_\mu$, up to $\sim 70$ GeV.
	
	Prospects for dark photon searches have been reported recently for several proposed collider and non-collider experiments and facilities. The low-mass range ($0.01-1$ GeV) is expected to be best covered by SHiP \cite{Anelli:2015pba}, NA62 in dump mode \cite{Lanfranchi:2018xrz}), and by FASER at the ATLAS interaction point \cite{Ariga:2018uku} in the very low-coupling regime ($\epsilon < 10^{-4}$). These are complemented by the LHCb Upgrade \cite{Ilten:2015hya,Ilten:2016tkc} and Belle-II \cite{Kou:2018nap}. Future collider experiments (HL-LHC \cite{Curtin:2014cca}, CEPC \cite{CEPCStudyGroup:2018ghi}, FCC-ee \cite{Karliner:2015tga}, FCC-hh \cite{Curtin:2014cca}, ILC500) have unique coverage in the high-mass range ($>$ 10 GeV) down to $\epsilon \sim 10^{-4}$. Very recently, Ref.~\cite{Tsai:2019mtm} updated the constraints of dark photon from NuCal and CHARM, and provided projections from the NA62, SeaQuest/DarkQuest, and LongQuest experiments.
	The so far unstudied prospects for dark photon searches at future electron-proton colliders may provide new opportunities as well as complementary coverage. 
	Compared to hadron colliders, the $ep$ counterparts are characterised by lower level of SM background, in particular from multi-jet production, allowing for larger signal-background ratio. The $ep$ colliders also avoid the problem of synchrotron radiation that occurs typically in circular lepton colliders. The main downside though is the relatively smaller scattering cross section.
	
	In this study, we focus on the Large Hadron electron Collider (LHeC) \cite{Dainton:2006wd,AbelleiraFernandez:2012cc,Klein:2018rhq} and the Future Circular Collider in hadron-electron collision mode (FCC-he) \cite{Mangano:2018mur,Benedikt:2018csr} in search of displaced vertex signatures of dark photons. The LHeC makes use of the 7-TeV proton beam of the LHC and a 60-GeV electron beam, to achieve a center-of-mass energy $\sim 1.3$ TeV with a total of 1 ab$^{-1}$ integrated luminosity, while the FCC-he would utilize the 50-TeV proton beam from the FCC resulting in a center-of-mass energy $\sim 3.5$ TeV and is expected to reach 3 ab$^{-1}$ integrated luminosity. Details on the proposed detector layout and expected performance can be found for instance in Ref.\ \cite{AbelleiraFernandez:2012cc}.

	\section{The model}
	\label{sec:model}
	We give a brief introduction to the theory of dark photons in this section. The dark photons here considered, cf.\ \cite{Okun:1982xi,Galison:1983pa,Holdom:1985ag,Boehm:2003hm,Pospelov:2008zw}, are light particles with mass in the MeV-GeV range and weakly coupled to the electrically charged SM particles. They are well motivated as portals to dark matter sectors, and correspond to an extension of the SM gauge group by an additional (broken) gauge group $U(1)_X$, with the associated gauge field $X_\mu$ coupled to the SM hypercharge gauge field $B_\mu$ through kinetic mixing. Equivalently speaking, the Lagrangian of the model includes a kinetic term proportional to $X_{\mu\nu} F^{\mu\nu}$ mixing the Abelian gauge bosons, where $X_{\mu\nu}$ and $F^{\mu\nu}$ are the field strength tensors of the $U(1)_X$ and the SM hypercharge gauge fields $X_\mu$ and $B_\mu$, respectively.
	
	After applying a field re-definition to get rid of the kinetic mixing term $X_{\mu\nu} F^{\mu\nu}$, we obtain the following term in the Lagrangian that gives rise to interactions between the dark photon field $A'$ and the SM fermion $f$,
	\begin{eqnarray}
		{\cal L} \supset - \sum_f \bar{f} \,\epsilon \, e \, q_f \, \fsl{A}'\, f,
		\label{eq:lagrangian}
	\end{eqnarray}
	with the electric charge $q_f$ of the fermion in the unit of $e$. As the additional $U(1)_X$ group is assumed to be broken (through some spontaneous breaking mechanism), the new gauge boson, dark photon labelled with $\gamma'$, is expected to be massive, i.e.~$m_{\gamma'}\neq 0$. The operators in Eq.~\eqref{eq:lagrangian} set the coupling strength between the dark photon and the SM fermions to be $\epsilon\,q_f\,e$.
	Depending on the mass of the dark photon, there are various decay modes into a pair of leptons, multiple hadrons, or quarks. The partial decay width of the dark photon into a single pair of charged leptons can be expressed with the following formula:
	\begin{eqnarray}
		\Gamma(\gamma' \rightarrow l^+ l^-) = \frac{1}{3}\, \alpha_{\text{QED}} \, m_{\gamma'} \, \epsilon^2 \sqrt{1-\frac{4 m_l^2}{m^2_{\gamma'}}} \bigg( 1 + \frac{2 m_l^2}{m^2_{\gamma'}} \bigg),
		\label{eq:decaywidth1}
	\end{eqnarray}
	where $\alpha_{\text{QED}}\sim 1/137$ is the QED fine structure constant and $m_l$ represents the mass of the lepton $l$ ($l=e, \mu, \tau$). We follow Ref.~\cite{Buschmann:2015awa} and compute the total decay width of the dark photon via its partial decay width to a pair of electrons and the branching ratio BR$(\gamma' \rightarrow e^- e^+)$:
	\begin{eqnarray}
		\Gamma_{\text{total}}(\gamma')=\frac{\Gamma(\gamma' \rightarrow e^- e^+)}{\text{BR}(\gamma' \rightarrow e^- e^+)}.
		\label{eq:decaywidth2}
	\end{eqnarray}
	We extract from Ref.~\cite{Raggi:2015yfk} the branching ratio of the dark photon into an electron pair as a function of the dark photon mass.
	
	In order to perform Monte-Carlo (MC) simulation, we employ the model file ``Hidden Abelian Higgs Model" provided in FeynRules \cite{Alloul:2013bka,Degrande:2011ua} model database, which was programmed according to the model description given in Ref.~\cite{Wells:2008xg}.

	\section{Analysis and results}
	\label{sec:results}
	\begin{figure}[t]
		\centering
		\includegraphics[width=0.6\textwidth]{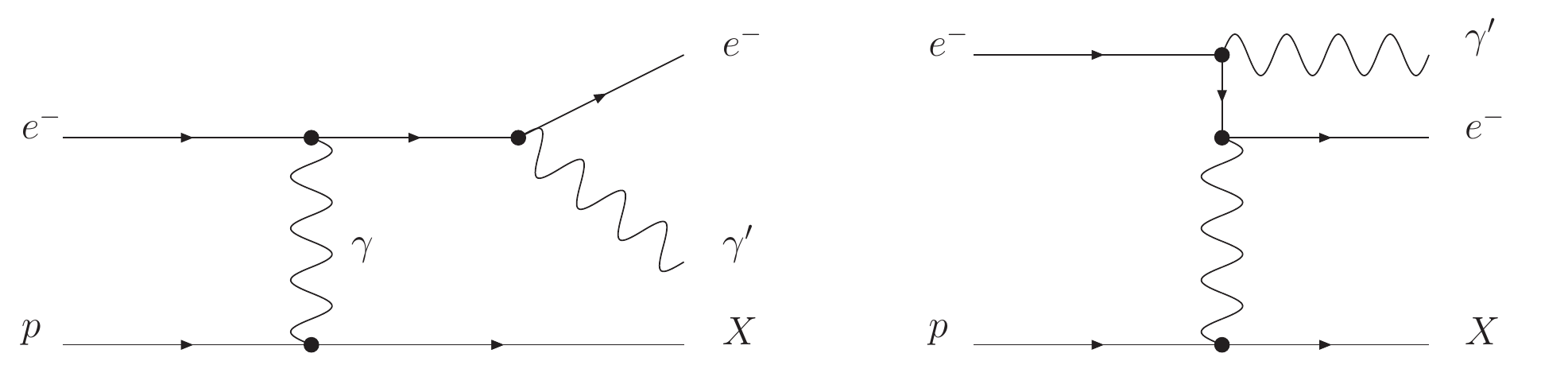}
		\caption{Feynman diagrams for the dark photon production processes in electron-proton collisions. Here $p$ and $X$ denotes a parton from the beam proton before and after the scattering process, respectively.}
		\label{fig:production}
	\end{figure}
	\begin{figure}[t]
		\centering
		\includegraphics[width=0.3\textwidth]{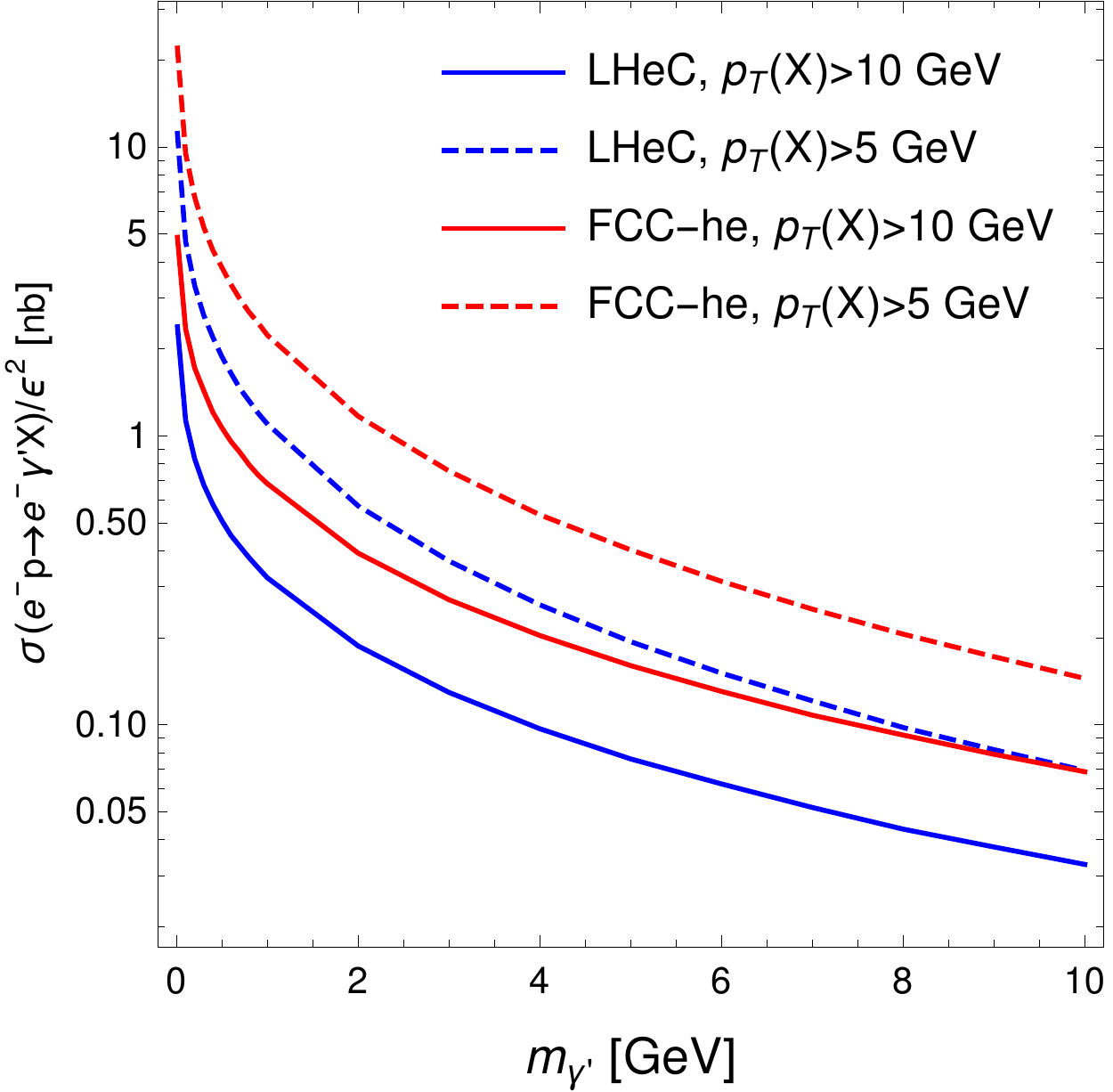}
		\caption{Production cross section for dark photons, via the process $e^- p \to e^- \gamma' X$, with $X$ denoting a number of hadrons. The dashed and solid line represents the lower transverse momentum cut on $X$ to be 5 and 10 GeV, respectively.}
		\label{fig:xsection}
	\end{figure}
	The dark photon production process in electron-proton collisions is shown via the corresponding Feynman diagrams in Fig.\ \ref{fig:production} while the production cross section, divided by $\epsilon^2$, as a function of $m_{\gamma'}$, is shown in Fig.\ \ref{fig:xsection}. The cross section is shown for two different transverse momentum cuts on the final state hadron, which puts a corresponding lower limit on the momentum transfer between electron and proton (labelled with $Q^2$).
	This is important as we limit ourselves to the deep inelastic scattering (DIS) regime, which means that the squared momentum transfer has to be much larger than the proton mass: $Q^2 \gg m_p^2 \simeq 1$ GeV$^2$.
	In practice we require a minimal transverse momentum of the final state parton of 5 GeV or 10 GeV, which sets a minimum value for $Q$. It is worthy of note that other production mechanisms exist, e.g.\ deep virtual Compton scattering (DVCS) and Bethe-Heitler-like processes (cf.\ e.g.\ \cite{Klein:2008di}) where the momentum transfer is sufficiently small to allow electron-proton instead of electron-parton scattering.
	The former (latter) process is expected to have a comparable (larger) cross section and results in larger (smaller) angles for the $\gamma'$ emission. We expect that these processes could potentially increase the signal strength. Nonetheless, a quantitative statement requires a dedicated analysis, which is beyond the scope of this paper.
	\begin{figure}
		\centering
		\includegraphics[width=0.4\textwidth]{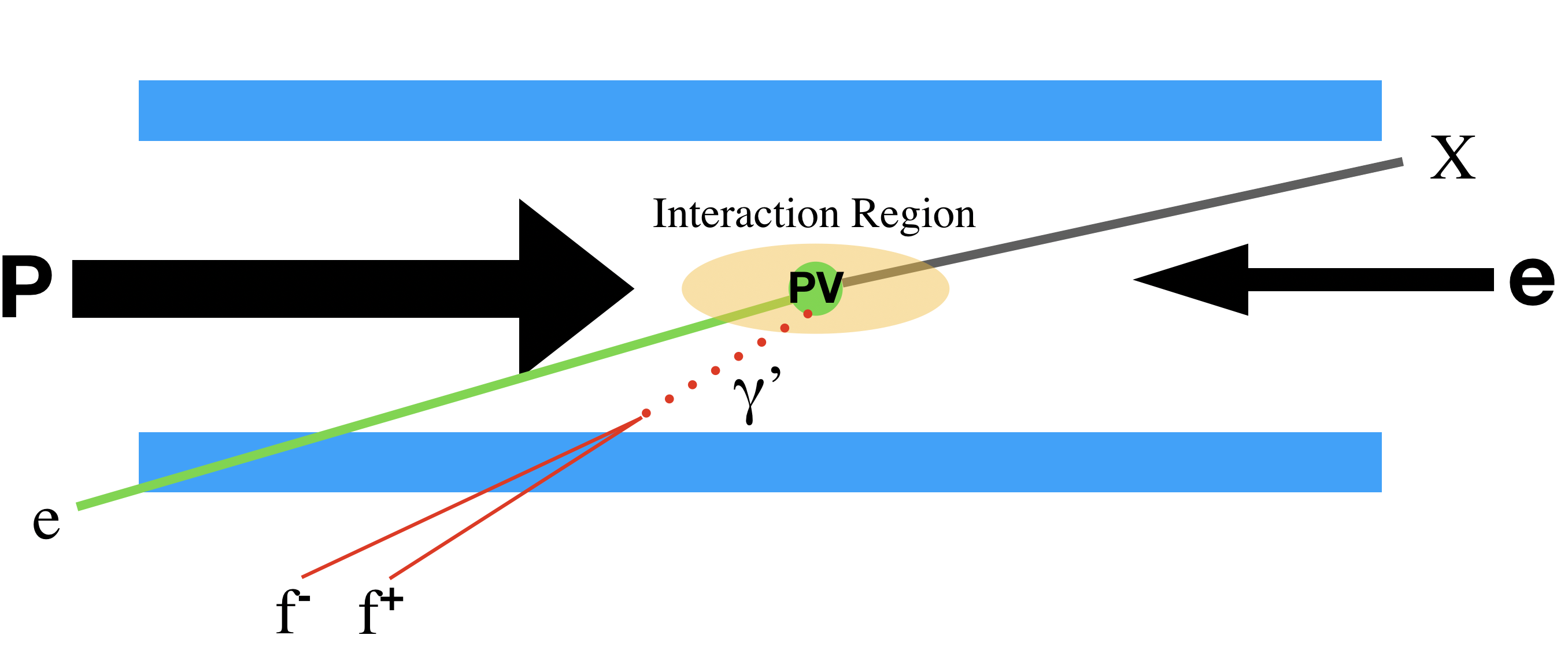}
		\caption{Sketch of the signal signature of a displaced dark photon decay. The proton (electron) beam is denoted by the larger (smaller) arrow from left to right (from right to left). The position of the primary vertex is inferred from the hadronic final state $X$ and the scattered electron $e$. From the primary vertex (labeled ``PV'' ) inside the interaction region the dark photon $\gamma'$ emerges and decays after some finite distance into the two charged particles $f^+$ and $f^-$.}
		\label{fig:ep_darkphoton_cartoon}
	\end{figure}

	The signal is given by the process $e^- p \to e^- X \gamma'$, where $X$ denotes the final state hadrons, and the dark photon $\gamma'$ decays into two charged fermions or mesons. This process is shown schematically in Fig.~\ref{fig:ep_darkphoton_cartoon}. In general in collisions with low momentum transfer the scattering angles of the electron and $X$ are small compared to the respective beams. Therefore the electron and proton beams are used to define the backward and forward hemispheres of the detector, which are optimized for low energy electromagnetic radiation and high energy hadrons, respectively.
	
	\begin{figure}
		\centering
		\includegraphics[height=0.25\textheight]{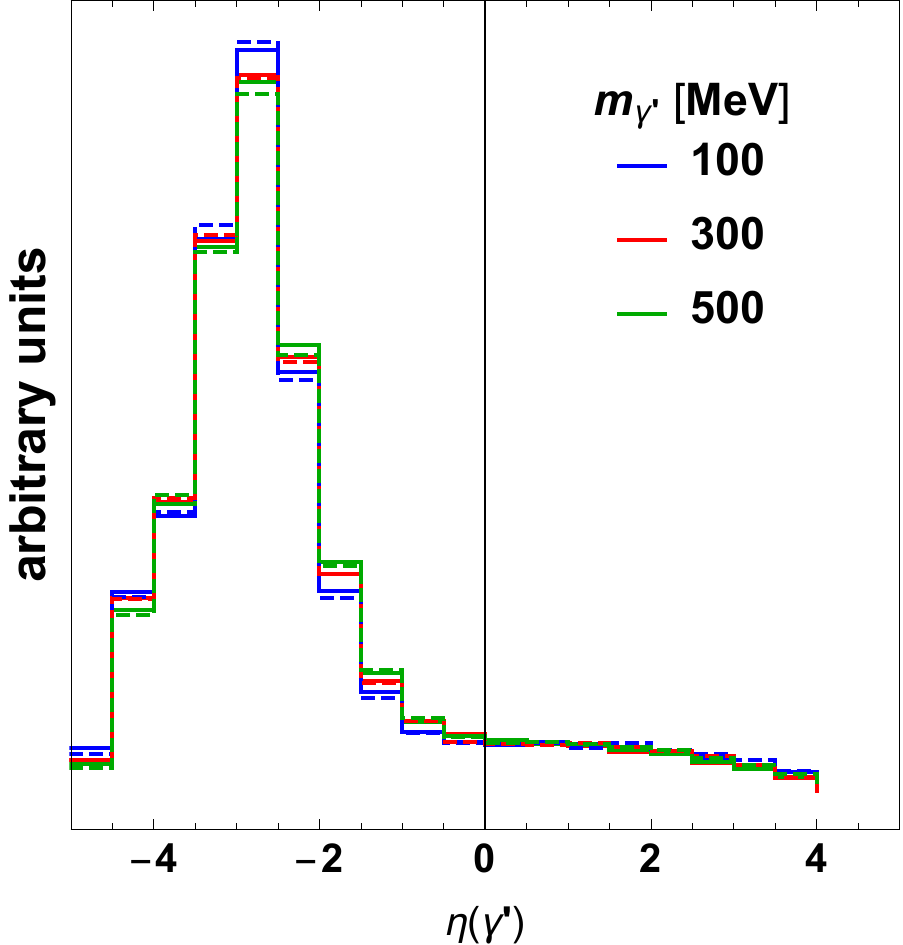}\qquad
		\includegraphics[height=0.25\textheight]{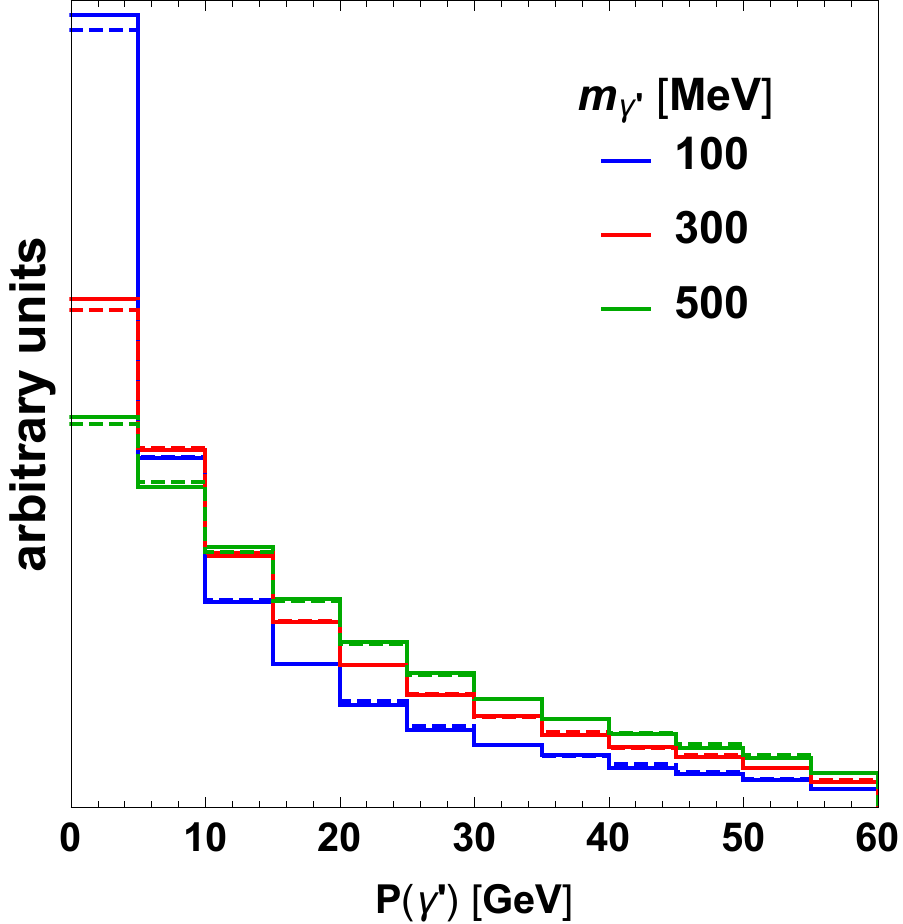}
		\caption{Dark photon distributions for eta (left) and momentum (right) at the LHeC (solid lines) and at the FCC-he (dashed lines). The colored lines are labelled according to the dark photon mass, $m_{\gamma'}$ in MeV. For this plot $P_T(X) > 5$ GeV is enforced. The scale is linear.}
		\label{fig:kinematics}
	\end{figure}
	Characteristic for the DIS production process of the dark photon are the small scattering angles of the deflected electron and parton from the beam interaction, which are, however, still within the geometric acceptance of the LHeC and FCC-he detectors. 
	The $\gamma'$ is typically emitted from the electron and has a very small emission angle.
	Exemplary for the $\eta$ and momentum distributions of the dark photon are shown for three different masses in fig.\ \ref{fig:kinematics}. Therein, the solid and dashed lines denote the LHeC and FCC-he, respectively. The typical acceptances are  $-4.3 < \eta < 4.9$ ($-5.0 < \eta < 5.2$) for electrons and muons, and $\eta$ up to 5 (5.5) for jets at the LHeC (FCC-eh). In practice, we implemented a cut on the pseudorapidity of each particle species as $|\eta|<4.7$. It is interesting to note that the difference for the two distributions between the two colliders is tiny; we interpret this as a consequence of the process to be more sensitive to the electron beam parameters and to prefer small momentum transfer.
	We find in our numerical simulation that the decay products, the fermion pair, carry a low momentum, and a transverse momentum that is roughly twice the dark photon mass.
	For $m_{\gamma'} > 10$ MeV, the resulting transverse momentum together with the magnetic field in the detector with $B=3.5$ T yields a gyroradius of order 10 cm for electrons, which is larger than the radius of the beam pipe\footnote{In order to perturb the proton beam as little as possible the electron beam is bent very strongly by the focusing magnets, which gives rise to a radiation fan that affects the symmetry of the beam pipe; its radius is 2.2 cm on three sides and 11 cm in direction of the radiation fan. Since the radiation does not give rise to secondary vertices, we do not take this asymmetry into account in our computation.} 
	The charged fermions and mesons from the dark photon decay are thus expected to enter the detector and spiral along the beam pipe close to the scattered electron.
	The flight length of the charged
	particle pair, given by distance between the secondary vertex and the backward calorimeter, allows for several cm of separation between the lepton and the anti-lepton.
	
	Possible backgrounds for dark photon masses below the muon production threshold could arise from real low-energy photons, produced for instance via Bethe-Heitler or DVCS processes, with the photon interacting with the detector material or the beam pipe. This could give rise to electron-positron pairs with similar kinematic properties to our process. 
	In this kind of background, however, the secondary vertex is expected to coincide with the known location of the detector material or the beam pipe, and we assume that they can be safely rejected on these grounds.
	Possible backgrounds also for larger dark photon masses might arise from long-lived mesons, such as $K_S$, $K_L$, and $\Lambda$ baryons with lifetimes of about 3 cm, 15 m, and 8 cm, respectively are expected to decay far away from the interaction point. Moreover, hadronic activity is aligned with the proton beam and propagates mostly into the forward hemisphere of the detector, and 
	their primary decay channels are only marginally consistent with our signal signature and their masses are well known, such that we expect that they can be vetoed efficiently without much loss of signal efficiency. 
	Cosmics or other machine related backgrounds are not expected to be relevant as long as they do not point to the primary vertex.
	
	As shown schematically in Fig.~\ref{fig:ep_darkphoton_cartoon} the primary vertex of the signal can be inferred from the scattered electron and the hadrons, and the secondary vertex from the spiraling charged particles.
	The projected experimental resolution for the vertexing is between $10\,\mu$m and $100\,\mu$m where the larger value is relevant for particle tracks with smaller energy and angles.
	For concreteness and to be conservative we assume that decays of the dark photon that have a displacement of at least $200\,\mu$m from the primary vertex can be detected and are free of background.
	The expected number of dark photon decays with a given displacement can be quantified with:
	\begin{align}
		&N_{\rm dv}(\sqrt{s},{\cal L},m_X,\epsilon) 
		= 
		\sigma(M,\epsilon)\, {\cal L} \times \int D(\vartheta,\gamma)\,P_{\rm dv}(x_{\rm min}(\vartheta),x_{\rm max}(\vartheta),\Delta x_{\rm lab}(\uptau,\gamma))\, d\vartheta d\gamma\,.
		\label{eq:masterequation}
	\end{align}
	In this equation, $\sigma$ labels the production cross section for dark photon at the $ep$ collider, and ${\cal L}$ denotes the integrated luminosity, $D(\vartheta,\gamma)$ is the probability distribution for $\gamma'$ with the Lorentz boost $\gamma$ and the angle $\vartheta$ between its momentum and the beam axis, $P_{\rm dv}$ is the probability distribution of a displaced decay, and $\Delta x_{\rm lab}$ denotes the mean decay length of the dark photon in the laboratory frame. 
	The proper lifetime $\uptau$ is obtained from the total decay width, which may be calculated with Eq.~\eqref{eq:decaywidth1} and Eq.~\eqref{eq:decaywidth2}.
	The probability of a displacement from the primary vertex with $x_{\rm min} \leq \Delta x_{\rm lab} \leq x_{\rm max}$ is given by
	\begin{equation}
		P_{\rm dv}= {\rm Exp}\left(\frac{-x_{\rm min}}{\Delta x_{\rm lab}}\right) -  {\rm Exp}\left(\frac{-x_{\rm max}}{\Delta x_{\rm lab}}\right)\,.
		\label{eq:probability}
	\end{equation}
	The dark photon lifetime is typically too small for its displaced decay to take place outside the beam pipe.
	We thus consider the displacements to be visible if they are larger than $x_{\rm min} = 200\,\mu$m, corresponding to the tracking resolution, and $x_{\rm max} = \infty$.
	In the laboratory frame the displacement is governed by the mean decay length:
	\begin{equation}
		\Delta x_{\rm lab} = \uptau_{\rm lab} |\vec v| = \sqrt{\gamma_{\gamma'}^2-1}\, \uptau c\,,
	\end{equation}
	with $c$ the speed of light and $\gamma_{\gamma'} = \sqrt{1+ |\vec p|^2/M^2}$ being the Lorentz factor with the three-momentum in the laboratory frame $\vec p$.
	Eq.~\eqref{eq:masterequation} thus allows us to compute the total number of dark photon decays that occur with a displacement from the primary vertex of at least $200\,\mu$m. 
	The probability $P_{\rm dv}$ for a dark photon to decay within these distances depends on the proper lifetime $\uptau$ of the dark photon and the corresponding Lorentz boost.

	We simulate the kinematics for the $\gamma'$ from the DIS production process with WHIZARD 2.6.4 \cite{Kilian:2007gr,Moretti:2001zz} using the built-in PDF sets, extracting the momentum and angular distributions. For the parton from the incoming proton beam as well as the outgoing hadrons we use up and down quarks and their antiparticles. This limitation on the light quarks leads to an underestimation of the production cross section, which also has contributions from strange and charm quarks. We estimate that the increased production cross section will be ${\cal O}(10)\%$ bigger, which renders our result somewhat conservative. 
	We consider the benchmark masses $m_{\gamma'}=10$ MeV and from 50 to 800 MeV in steps of 50 MeV.
	
\begin{figure}
		\centering
		\includegraphics[width=0.4\textwidth]{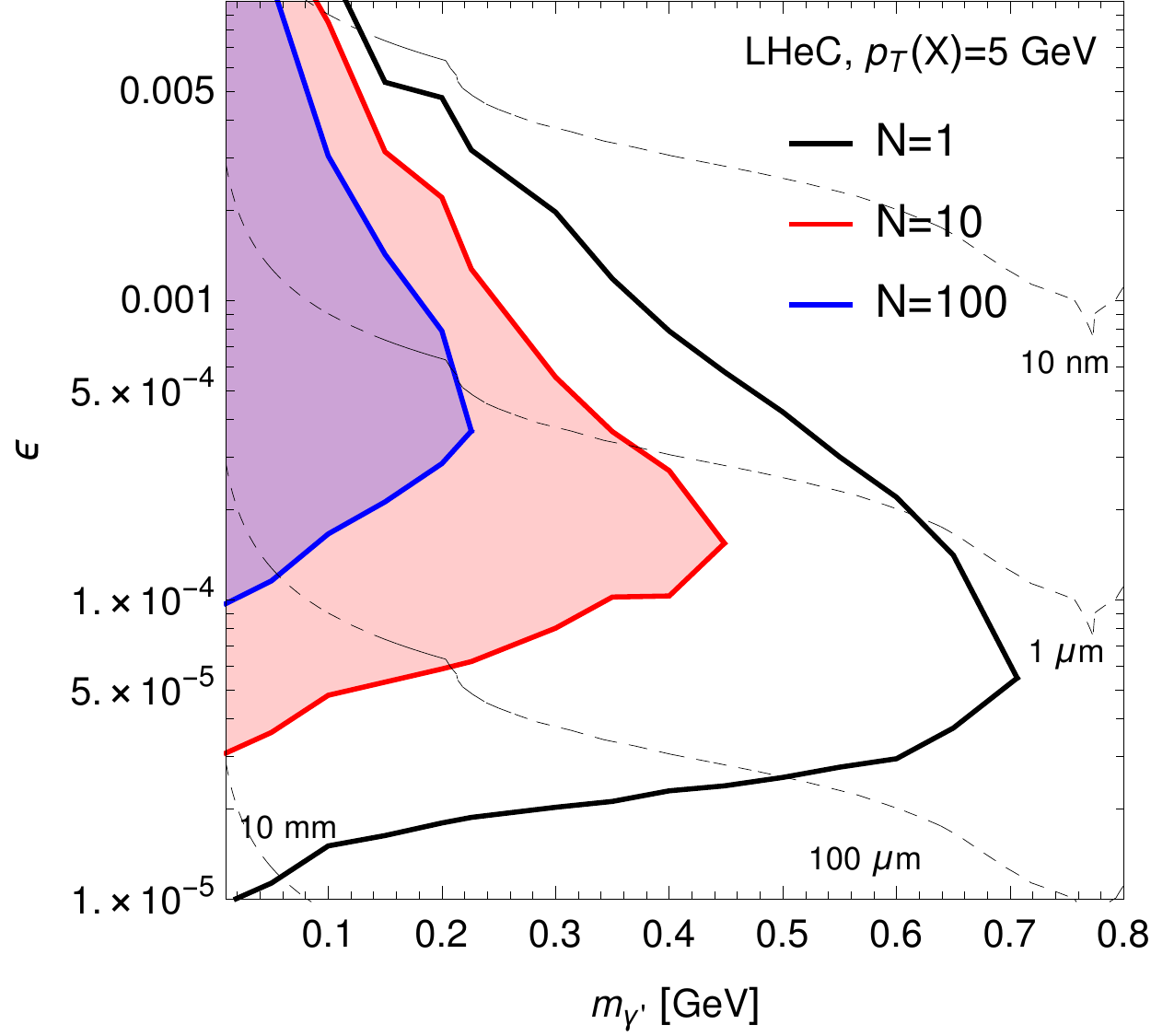}
		\includegraphics[width=0.4\textwidth]{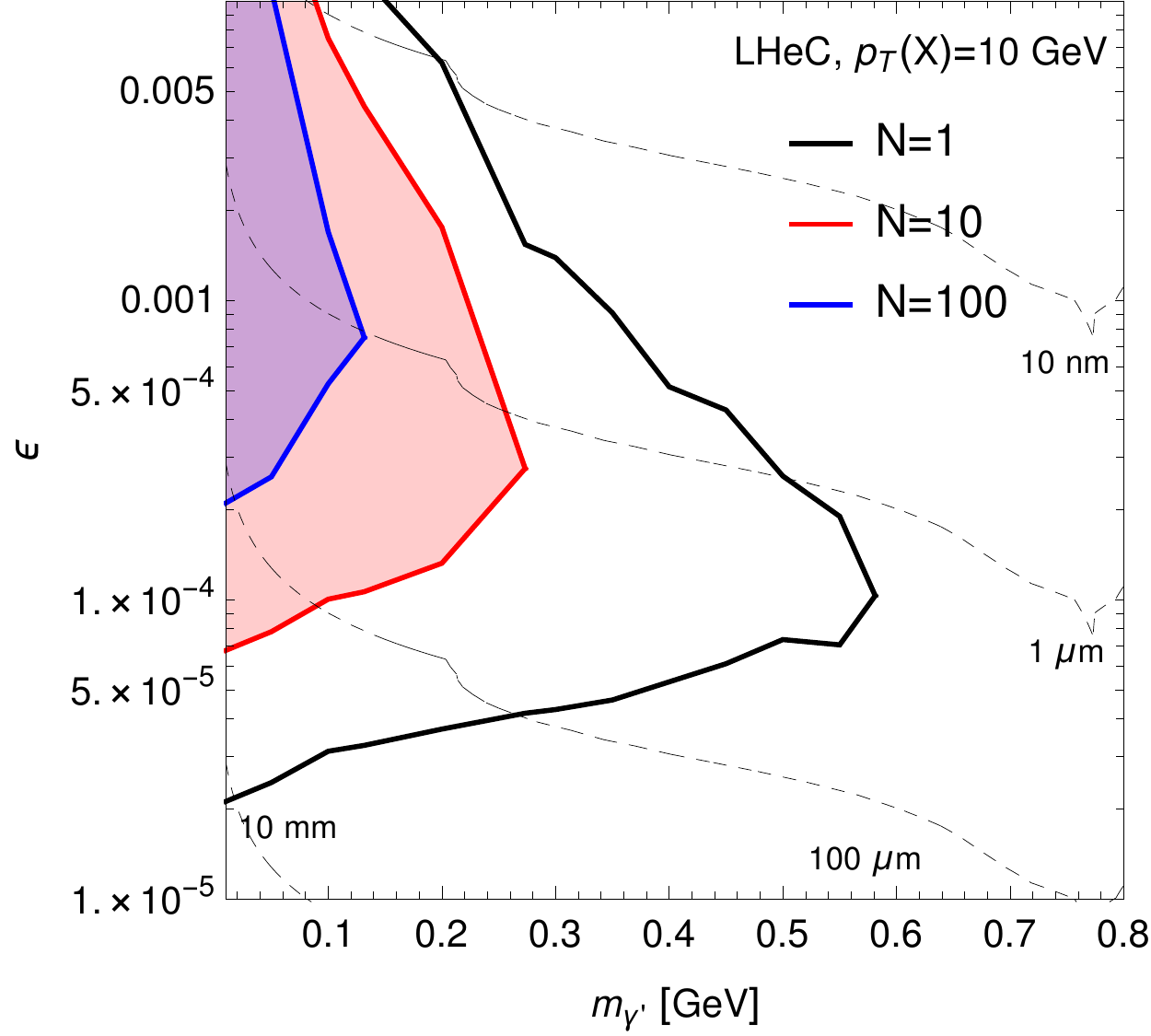}
		
		\includegraphics[width=0.4\textwidth]{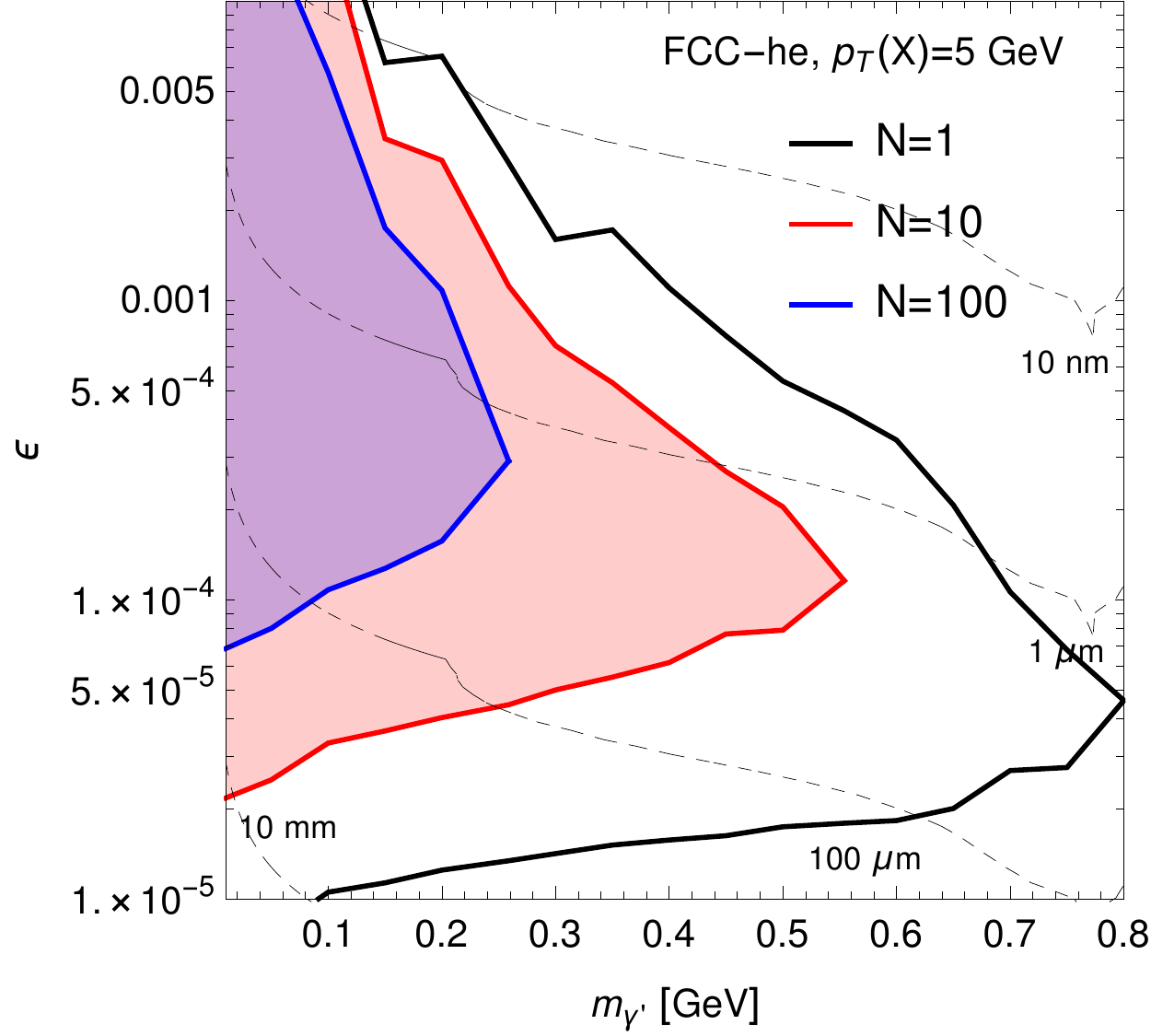}
		\includegraphics[width=0.4\textwidth]{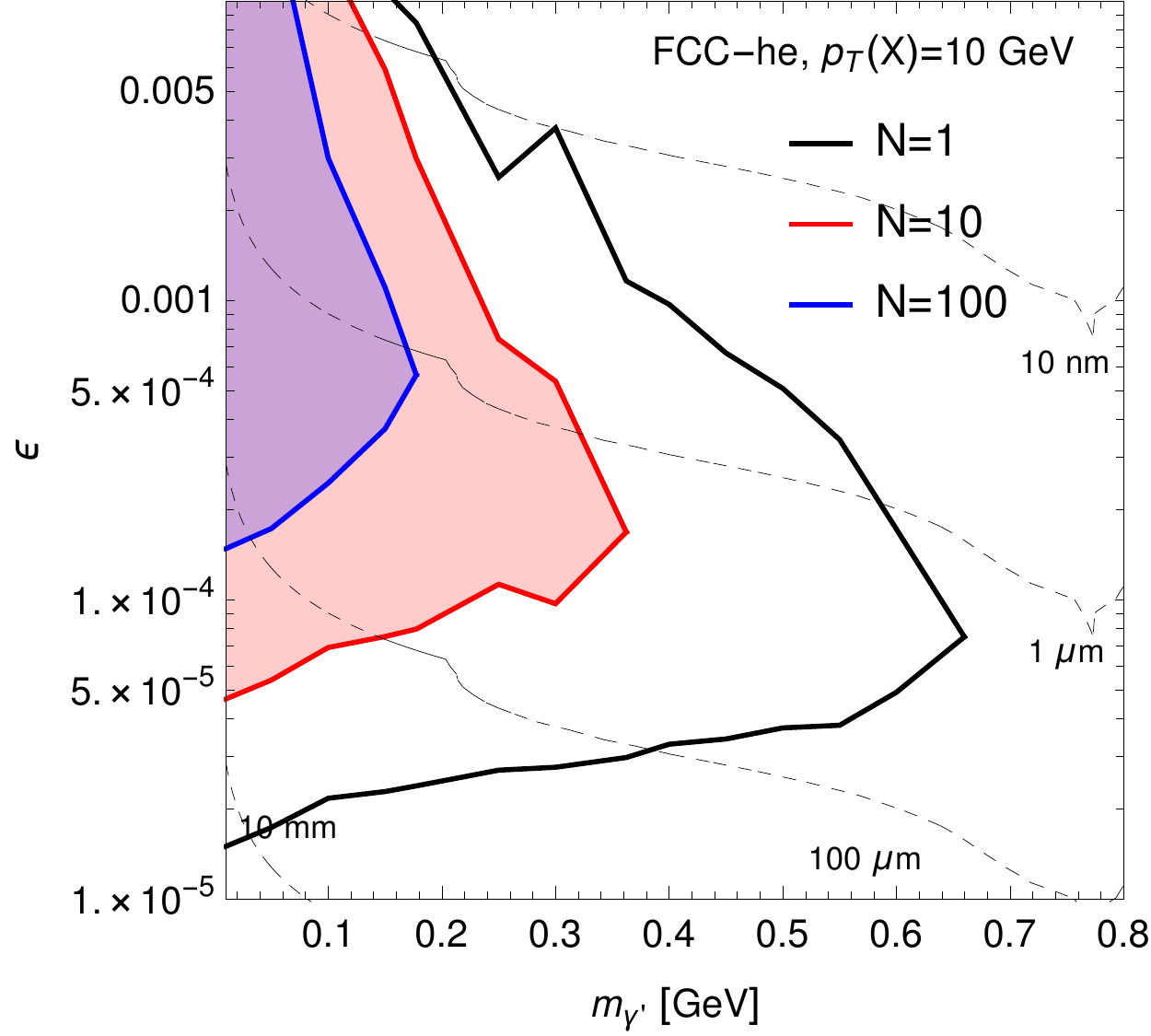}
		\caption{Parameter space contours for a number $N=1, 10, 100$ expected dark photon decays at the LHeC (top row) and the FCC-he (bottom row) for final state hadron $P_T$ cuts of 5 GeV (left column) and 10 GeV (right column). The black dashed lines denote isocontours for selected proper lifetimes. In this figure, zero background and 100\% signal efficiency are assumed.}
		\label{fig:contours}
	\end{figure}

	First, we assume that a displaced decay of a dark photon into a pair of charged SM particles can be detected with 100\% efficiency and that all the above mentioned background processes are reducible without further effect on the signal efficiency.
	We show the contour lines for N=1, 10, and 100 expected dark photon decays at the LHeC and the FCC-he under this assumption in the four panels of Fig.\ \ref{fig:contours}.
	
\begin{figure}[t]
		\centering
		\includegraphics[height=0.23\textheight]{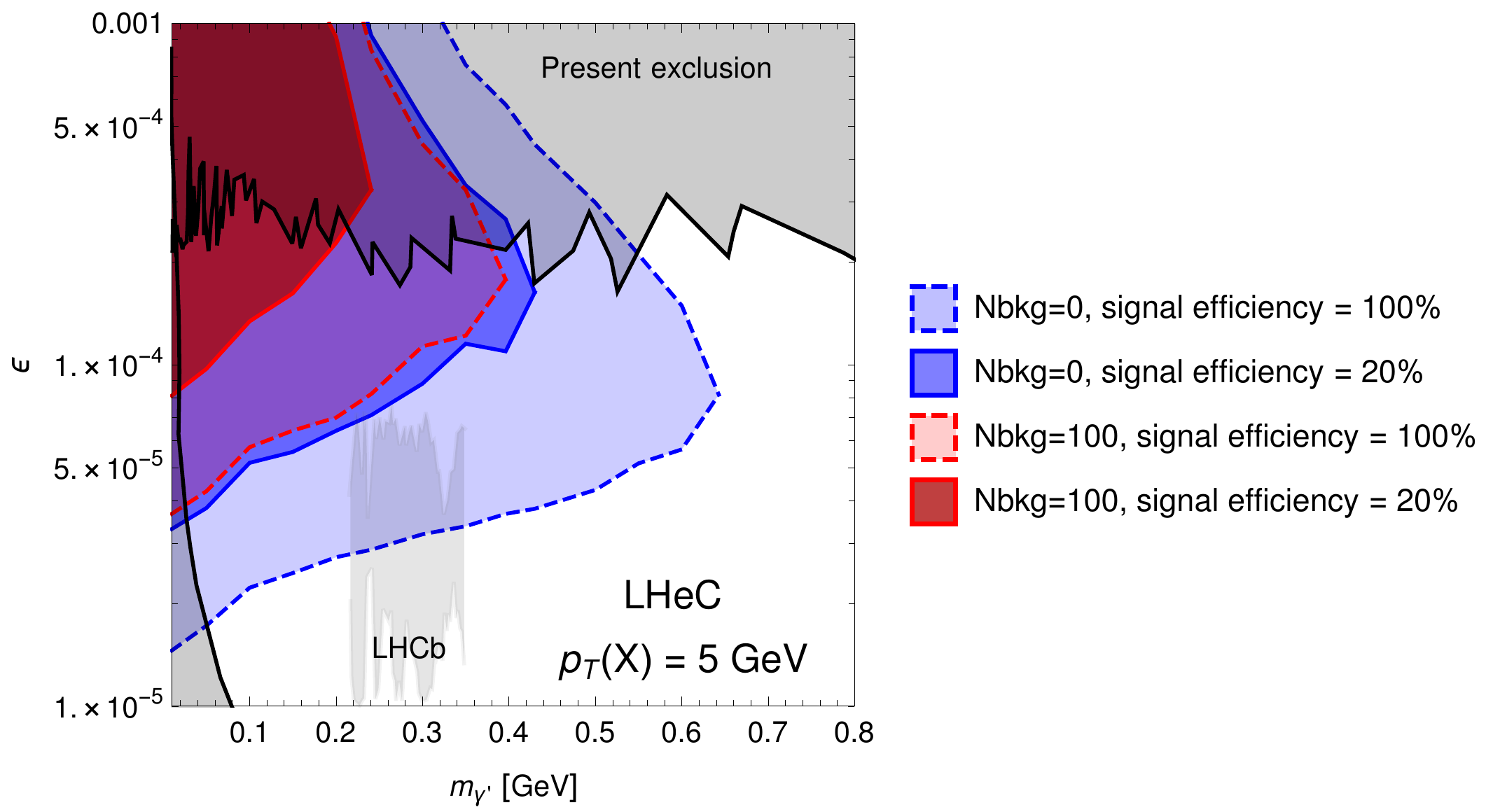}
		\includegraphics[height=0.23\textheight]{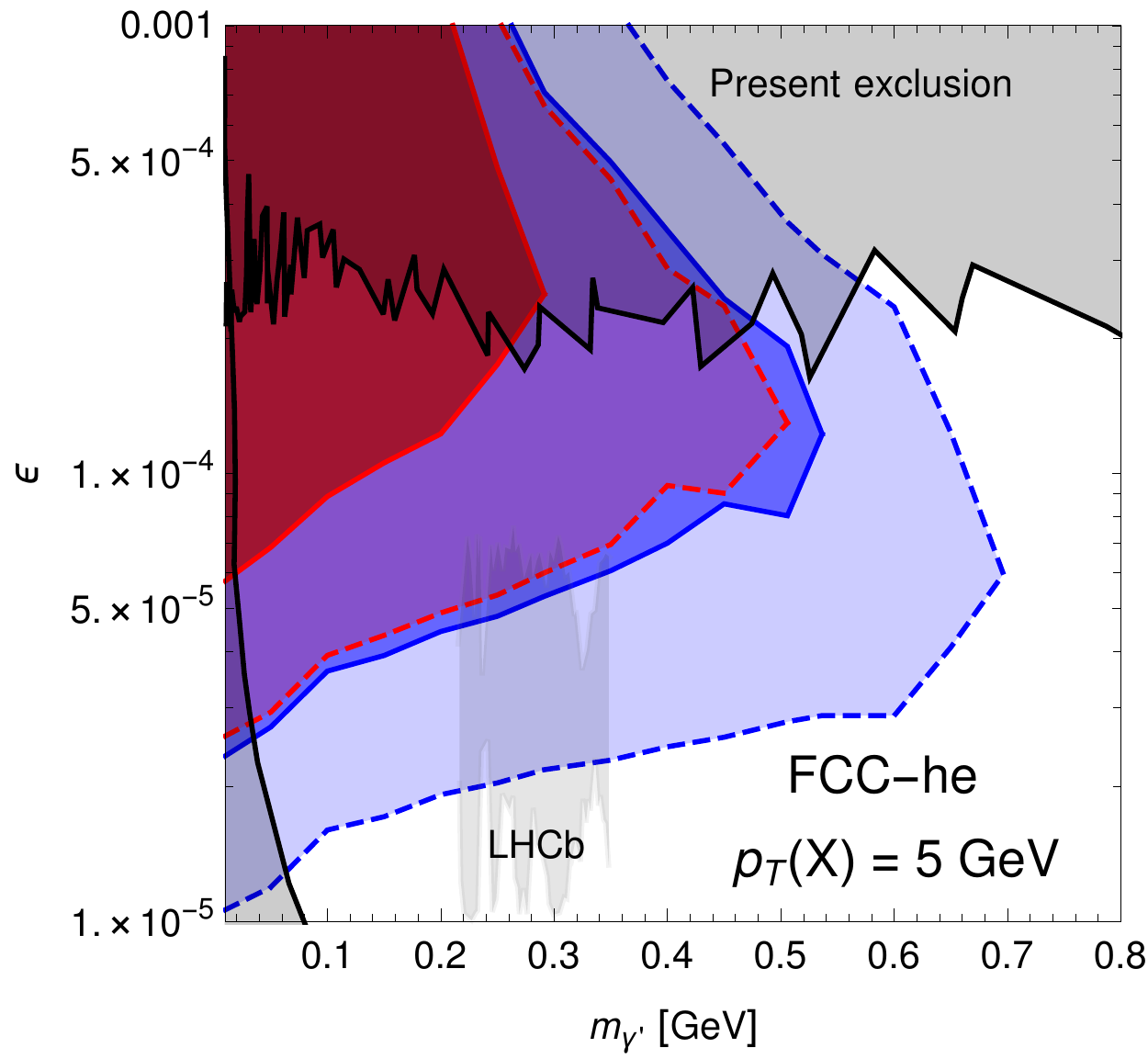}
		\caption{Projected sensitivity of dark photon searches at the LHeC and FCC-he via displaced dark photon decays. The sensitivity contour lines are at the 90\% confidence level and consider a transverse momentum cut on the final state hadrons of 5 GeV. The blue and red areas denote the assumption of zero and 100 background events, respectively, the solid and dashed lines correspond to a reconstruction efficiency of 100\% and 20\%, respectively. The shaded gray area labelled with ``LHCb'' is currently being tested by the LHCb experiment \cite{Aaij:2017rft}. See text for details.}
		\label{fig:darkphoton_sensitivity}
	\end{figure}

	The above assumption on number of background events and signal efficiency is an optimistic approximation, and in a real experiment irreducible backgrounds may exist, the rejection of which, along with reconstruction losses and further detector effects, may affect the sensitivity of the experiment to the dark photon signature.
	To get an impression on how these effects modify our prediction for the exclusion power of the LHeC and the FCC-he, we show contour lines for four different hypotheses at the 90\% confidence level (CL) in Fig.~\ref{fig:darkphoton_sensitivity}, for the LHeC and the FCC-he with a total integrated luminosity of 1 ab$^{-1}$ and 3 ab$^{-1}$, respectively.
	For the signal significance at 90\% CL with zero background events we require 2.3 and 11.5 events for signal and triggering efficiencies of 100\% and 20\%, respectively; similarly for 100 background events we require 14.1 and 70.4 signal events for efficiencies of 100\% and 20\%, respectively.
	The final exclusion sensitivity of the LHeC and the FCC-he for the considered number of background events is inside the colored area, depending on the real signal efficiency.
	
	In Fig.~\ref{fig:darkphoton_sensitivity} we consider final state hadrons with transverse momentum of at least 5 GeV to ensure the DIS regime of the production process.	
	Although being very small we do not expect the momentum threshold of 5 GeV to pose a problem to the experimental analysis, since the final states (consisting of an electron with about 60~GeV, hadrons in forward direction, and two low-energy leptons or mesons in backward direction) all typically have scattering angles of a few degrees with respect to the beams, which are well within the geometric detector acceptance.
	
	Also included in the figure are the present exclusion bounds on the dark photon, denoted by the gray area. The limits in the lower left corner of the figure stem from dark photon searches at the beam dump experiments E141 \cite{Riordan:1987aw}, E774 \cite{Bross:1989mp}, one in Orsay \cite{Davier:1989wz}, and the updated constraints from the NuCal experiment from Ref.\ \cite{Tsai:2019mtm}; the upper limits on the mixing from the beam dump experiment NA48 \cite{Batley:2015lha} and the electron-positron collider experiment BaBar \cite{Lees:2014xha}. Shown by the light gray region and labelled with ``LHCb'' is the region currently tested by the LHCb experiment in their search for long-lived particles \cite{Aaij:2017rft}.
	A preliminary evaluation of the sensitivity to dark photons at the LHeC and FCC-he as presented in this paper had been reported in Ref.~\cite{Strategy:2019vxc} (Fig.~8.16), wherein they are compared with the potential sensitivity of several other future facilities, illustrating how electron-proton colliders offer a complementary coverage in a low-mass and intermediate coupling regime.
	
	It is important to realize that in particular the final state electrons are very difficult to test in any other present and future experiment for masses below the di-muon production threshold.
	Electron-proton colliders will offer an excellent coverage for dark photon masses around 0.2 GeV and mixing above $10^{-5}$.
	
	\section{Conclusions}
	\label{sec:conclusions}
	Extending the SM gauge group with an additional $U(1)_X$ factor gives rise to a dark photon that interacts with the SM fermions via kinetic mixing.
	The interaction strength is governed by the mixing parameter $\epsilon$, which also leads to dark photon decays into pairs of leptons, hadrons, or quarks.
	In this article we have estimated the prospect for a dark photon search at the LHeC and FCC-he via its displaced decays into two charged SM particles and for a mass range 10 MeV $\leq m_{\gamma'}\leq$ 0.7 GeV.
	Under the assumption that unknown backgrounds are completely reducible and can be suppressed without much loss of signal efficiency, we found that non-observation of a signal at the LHeC (FCC-he) can exclude dark photons in the considered mass range with kinetic mixing $\epsilon$ larger than about $2\times 10^{-5}$ ($10^{-5}$) when considering final state hadrons with transverse momentum above 5 GeV. 
	This complements existing searches and search strategies for dark photons in this mass range, which usually probe mixings either much below these numbers, or above~$10^{-3}$.
	It also complements forecasted sensitivities at future colliders, which cover mostly the large-mass, large-coupling regime, and also the low-mass, very low-coupling sensitivity of beam-dump or fixed-target experiments, or external LHC detectors such as FASER. 
	The electron-proton colliders would therefore offer a complementary coverage in a low-mass and intermediate coupling regime.

	\section*{Acknowledgements}
	We thank Florian Domingo, Herbi Dreiner, Ahmed Hammad, and Max Klein for useful discussions. Z.~S.~W. is supported by the Sino-German DFG grant SFB CRC 110 ``Symmetries and the Emergence of Structure in QCD", the Ministry of Science, ICT \& Future Planning of Korea, the Pohang City Government, and the Gyeongsangbuk-do Provincial Government through the Young Scientist Training Asia-Pacific Economic Cooperation program of APCTP. O.~F. received funding from the European Unions Horizon 2020 research and innovation program under the Marie Sklodowska-Curie grant agreement No 674896 (Elusives). M.~D. acknowledges the support of STFC, United Kingdom.
	

\end{document}